\documentclass[useAMS,usenatbib]{mn2e}

\usepackage{gensymb}
\usepackage{graphicx}
\usepackage{amssymb}
\usepackage{epsfig}
\usepackage{natbib}
\usepackage{epsf,palatino,url,graphicx,array,setspace,amssymb,fancyhdr,lscape,rotating,amsmath,xcolor}
\newcommand{\WHz}{\>{\rm W}\,{\rm Hz}^{-1}}

\title[Constraints on the AGN radio luminosity functions at low power.]{
  Looking below the floor: constraints on the AGN radio luminosity functions
  at low power.}  \author[Alessandro Capetti and Claudia
  M. Raiteri]{Alessandro Capetti$^{1}$\thanks{E-mail: capetti@oato.inaf.it
    (AC); raiteri@oato.inaf.it (CMR)} and Claudia
  M. Raiteri$^{1}$\\ $^{1}$INAF-Osservatorio Astrofisico di Torino, via
  Osservatorio 20, 10025 Pino Torinese, Italy}
\begin{document}

\date{}

\maketitle

\label{firstpage}

\begin{abstract}
We constrain the behavior of the radio luminosity function (RLF) of two
classes of active galactic nuclei (AGN) namely AGN of low radio power
  (LRP) and BL Lac objects. The extrapolation of the observed steep RLFs to
low power predicts a space density of such objects that exceeds that of the
sources that can harbor them and this requires a break to a shallower
slope. For LRP AGN we obtain $P_{\rm br,LRP} \ga 10^{20.5} \, \rm W \,
Hz^{-1}$ at 1.4 GHz to limit their density to be smaller than that of
elliptical galaxies with black hole masses $M_{\rm BH} > 10^{7.5} M_\odot$. By
combining this value with the limit derived by the observations the break must
occur at $P_{\rm br,LRP} \sim 10^{20.5} - 10^{21.5} \, \rm W \, Hz^{-1}$. For
BL~Lacs we find $P_{\rm br,BLLAC} \ga 10^{23.3} \, \rm W \, Hz^{-1}$ otherwise
they would outnumber the density of weak-lined and compact radio sources,
while the observations indicate $P_{\rm br,BLLAC} \la 10^{24.5} \, \rm W \,
Hz^{-1}$. In the framework of the AGN unified model a low luminosity break in
the RLF of LRP AGN must correspond to a break in the RLF of BL~Lacs. The ratio
between $P_{\rm br,LRP}$ and $P_{\rm br,BLLAC}$ is $\sim 10^3$, as expected
for a jet Doppler factor of $\sim 10$.
\end{abstract}

\begin{keywords}
galaxies: active -- galaxies: BL Lacertae objects: general --
     galaxies: jets
\end{keywords}

\section{Introduction}
Many astrophysical quantities obey a distribution well described by
power-laws. Important insights into the physical processes producing such
distributions can be obtained from their power-law index. In addition, the
location at which the distribution departs from a power-law is also of great
importance. In most cases, a change in the distribution laws is observed at
their high end, in the form of either a steepening of the distribution, or an
exponential cutoff. For example, the peak frequency in the synchrotron
emission of active galaxies is associated to the largest energy attainable by
relativistic electrons and it carries essential constraints on the
acceleration and cooling mechanisms. Equally important is the determination of
the behavior of the distributions at the low end as it might reveal the
presence of, e.g., a minimum mass for the formation of a star or a
galaxy. Unfortunately, the low end behavior is often unaccessible to
observations or, in other cases, it is blurred by the emergence of strong
selection biases.

In this Letter we focus on the radio luminosity function (RLF) of two classes
of active galaxies at low redshift, $z<0.1$. The first is formed by low radio
power (LRP) AGN: within the considered redshift limit, most of the radio
emitting AGN are indeed objects of low power, $\log (P_{r} / \WHz) \lesssim
24$.  The second consists of BL~Lac objects. According to the unified model
for AGN, these two RLFs are expected to be connected with each other, as LRP
AGN should represent the parent population of BL~Lac objects. In both cases
the RLFs are observationally well defined over a very broad range of radio
power. The RLF of LRP AGN also presents a break at high
luminosity. Conversely, at their low luminosity end, both RLFs are well
described by a pure power law by current observations, without clear signs of
a change in their slopes. In this Letter we show that it is nonetheless
possible to constrain the power at which a low luminosity break in the RLF
must occur.

Throughout the paper we adopt $H_0=70 \, \rm km \, s^{-1} \, Mpc^{-1}$, 
$\Omega_{\rm M}=0.30$, and $\Omega_\Lambda=0.7$

\section{The radio AGN luminosity function}
 
The number density of AGN cannot exceed that of their potential hosts:
this argument has been used by \citet{mauch07} and \citet{cattaneo09} to
constrain the low luminosity behavior of their RLF. They found that a break
must occur at a radio luminosity $\log (P_{r} / \WHz) \sim 19.5$ and $19.2$,
respectively. We use the same rationale, but including the information that
their hosts, as shown in more detail below, are associated almost exclusively
to very large black holes.

The RLF of nearby galaxies extending to the lowest luminosity
has been obtained by \citet{mauch07}. It is derived from a sample of 7824
radio sources from the 1.4 GHz NVSS survey cross-correlated with the second
data release of the 6 degree Field Galaxy Survey (6dFGS, \citealt{jones05}).
\citeauthor{mauch07} separated radio sources associated to AGN from
star-forming galaxies relying on their optical spectra. Their resulting RLF
covers the range $\log (P_{r} / \WHz)  = 20.4 - 26.4$ and shows
a steepening at high luminosities (above $\log (P_{r} / \WHz)  = 24.59$).

\citet{best12} obtained an independent estimate of the AGN RLF.  This is
  slightly shallower (reaching a luminosity $\log (P_{r} / \WHz) \sim 22$) but,
  since their sample is drawn from the SDSS, they could take advantage of the
  optical data to better characterize their properties. The main result is
  that their hosts are almost exclusively massive early-type galaxies with
  black hole masses (derived from the stellar velocity dispersion adopting the
  \citealt{tremaine02} law) larger than $\log \,(M_{\rm
    BH}/M_\odot)$$\sim$$7.8$ (see also \citealt{baldi10b}).

The SDSS survey provides us with a highly complete sample that can be used to
estimate the number density of galaxies with large black holes. To
this purpose, we analyze the 818,333 galaxies (MPA-JHU sample hereafter)
in the value-added spectroscopic catalog produced by the Max Planck Institute
for Astrophysics, and the Johns Hopkins University, and available at {\tt
  http://www.mpa-garching.mpg.de/SDSS/} \citep{bri04,tre04}. We consider all
galaxies from redshift $z=0.05$ to $z = 0.1$ to ensure a high level of
completeness (see below) within the largest possible volume, which results in
0.053 Gpc$^3$.

We then select elliptical
galaxies in the MPA-JHU sample by setting a threshold to the concentration
index $C_{r} \ge 2.6$ (e.g.\ \citealt{strateva01,kauffmann03b,bell03}) and to
their stellar velocity dispersion adopting, conservatively, $\sigma_{\rm star}
\ge 140 {\,\rm km\,s}^{-1}$, corresponding to $\log \,(M_{\rm BH}/M_\odot) \ge
7.5$, finding $\sim$38,000 objects. The resulting sample has a high level of
completeness. According to \citet{montero09} the completeness of the
SDSS decreases with decreasing apparent magnitude, starting at $\sim$ 95\% at
the SDSS spectroscopic limit of $r=17.77$, and being still higher than 80\% at
$r=13.25$. The vast majority of the LRP AGN hosts have a magnitude in the range
-23.5 $< M_K < $ -26.5 \citep{mauch07}. This translates into $r=13.3$--16.3 at
$z=0.05$ (and $r=14.8$--17.8 at $z=0.1$) having adopted $r-K=3.0$
\citep{chang06}. This implies that most of such galaxies in the selected
redshift range are included in the MPA-JHU catalog.

\begin{figure*}
   \includegraphics[width=85mm]{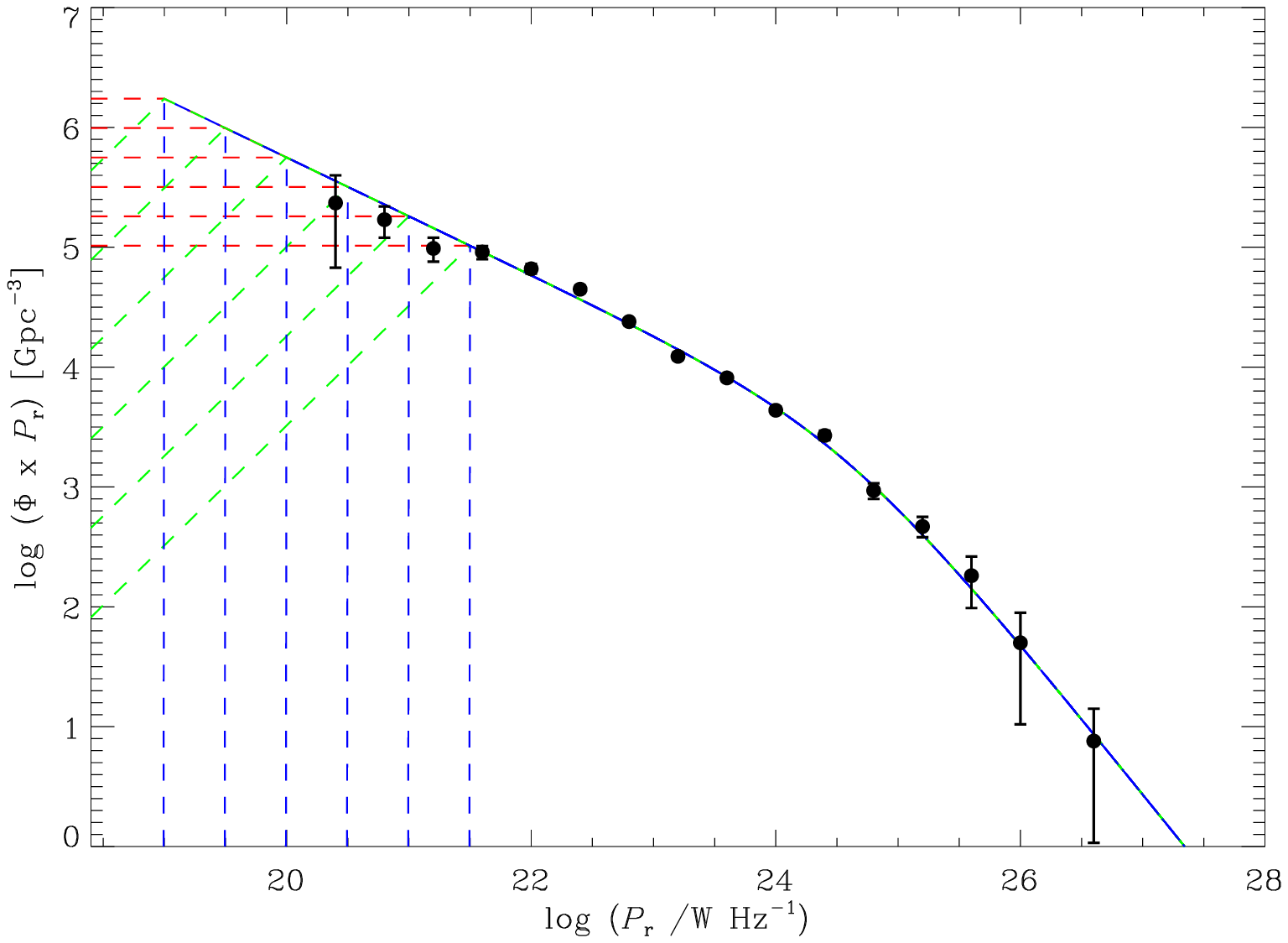}
   \includegraphics[width=87mm]{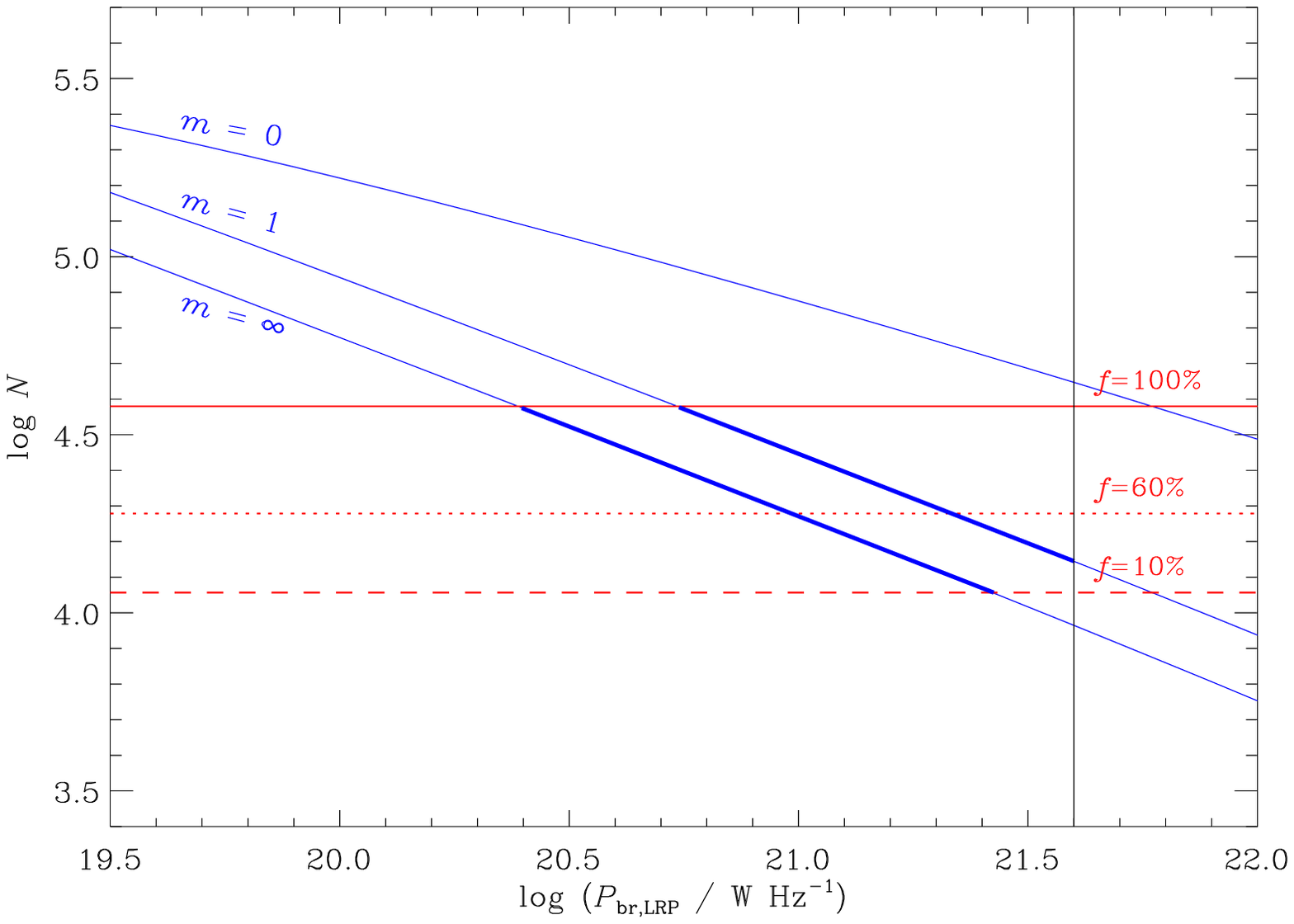}
   \caption{Left panel: AGN number density as a function of the radio power
     $P_{\rm r}$ at 1.4 GHz. The results of \citet{mauch07} are shown as black
     dots and their fit as a solid blue line. The fit has been extrapolated
     below different values of the break luminosity, $P_{\rm br,LRP}$, with
     power-laws of different slopes $m$ (red for $m=0$, green for $m=1$, and
     blue for $m=\infty$). Right panel: total number of predicted LRP AGN in the
     SDSS-DR7 area with $0.05 \le z \le 0.1$ as a function of the
     low luminosity break in the RLF, estimated for the three values of
     $m$. The solid horizontal line represents the total number of possible
     radio-galaxies hosts in the same volume; the other horizontal lines allow
     for smaller occupation fractions. The thick blue segments highlight the
     ranges of acceptable $P_{\rm br,LRP}$ values. The vertical line marks
     the upper limit to $P_{\rm br,LRP}$ set by the observations.}
\label{rge}
    \end{figure*}

We now estimate how the total number of predicted LRP AGN varies depending on
the behavior of the RLF at low power. A low luminosity break must occur
  in order to avoid the divergence of the LRP AGN number. The
  \citeauthor{mauch07} results suggest that it is located below 
$\log (P_{r} / \WHz)  \sim 21.6$, where the limited number of observed
  objects makes the RLF shape uncertain. We assume that the LRP AGN number
density follows, below a break luminosity $P_{\rm br,LRP}$, a power law with
an index $m$.  We investigate the effects of varying $m$ from 0 (a flat number
count distribution) to infinity (equivalent to a sharp cutoff below $P_{\rm
  br,LRP}$), see the left panel of Fig. \ref{rge}. We integrate numerically
the RLF starting from $\log (P_{r} / \WHz) \sim 18.4$, a factor of 100 below the
observational limit: the total number of predicted LRP AGN grows rapidly at
decreasing $P_{\rm br,LRP}$ (Fig.\ \ref{rge}, right panel). For $m=0$ it
exceeds the total number of available hosts when $\log (P_{\rm br,LRP} / \WHz)\sim
21.8$, a value that is inconsistent with the
\citeauthor{mauch07} results even assuming that {\sl all} massive galaxies
host a LRP AGN. For larger $m$ (and already for $m=1$) the tension with the
observations is eased, as we obtain $\log (P_{\rm br,LRP} / \WHz)\sim 20.4$--$20.7$. However, this requires an occupation fraction $f$ of
100\%, i.e., all massive galaxies must harbor a radio-source with at least
$P_{\rm r} \sim P_{\rm br,LRP}$. The bivariate RLFs, obtained by
\citeauthor{mauch07} splitting the sample into bins of infrared luminosity
(their Fig. 16), indicate instead values in the range $f \sim 10\% - 60 \%$
depending on the host magnitude. This implies that the values of $P_{\rm
  br,LRP}$ derived above should be considered as strict lower limits.

We conclude that: 1) the RLF of LRP AGN must break to a shallower slope for
radio luminosities smaller than $\log (P_{r} / \WHz)  \sim 20.5 - 21.5$, and 2) the number density below $P_{\rm br.LRP}$ must
significantly decrease, leading to a peak in its distribution.

\section{The BL~Lac objects luminosity function}

In this section we consider the RLF of BL~Lacs at 1.4 GHz. We adopt
the same line of reasoning followed above. In this case, the constraint on the
RLF is derived imposing the requirement that the total number of BL~Lacs does
not exceed the total number of appropriate (see below) radio sources in the
same volume.

\citet{best12} provides us with a sample of 18,286 AGN obtained by
combining the SDSS DR7 MPA-JHU sample with the NVSS and the FIRST surveys,
down to a flux density level of 5 mJy in the NVSS. We consider the same volume
as in Sect.~ 2, i.e., $0.05 \le z \le 0.1$. We select the sources with 1) a
rest-frame equivalent width of all emission lines smaller than 5 \AA\ (the
standard threshold for a BL~Lac classification, \citealt{stickel91}), and 2) a
flux of the central radio component in the FIRST images larger than 5 mJy.
The latter requirement (based on the measurement of \citealt{best12}) is
  needed to isolate potential BL~Lac objects according to their radio core
  emission while excluding, e.g., double lobed sources lacking of a core. The
  threshold is set at the same radio flux limit that defines the selected
  sample and it is included self-consistently in the analysis that follows.
We are left with 524 potential BL~Lac objects. We further check their radio
morphology: indeed BL~Lacs are only associated with specific radio structures
on the kpc scale, being dominated by a compact radio core and, in some cases,
showing also a one-sided jet or a halo \citep{antonucci85}. We found that 49\%
of the selected sources are instead well defined Fanaroff-Riley type I or II
(FR~I or FR~II) radio galaxies with sizes larger than 10--20 kpc. The 267
remaining objects are not necessarily all BL~Lacs and certainly include
intruders, for example, FR~I with size smaller than $\sim$10 kpc that, due to
the limited resolution of the FIRST images, are not recognized as such.
However, with only a handful of exceptions, the selected objects are all
massive early-type galaxies, conforming with the standard LRP AGN hosts.

On the other hand, there is the possibility that BL~Lacs are so nuclearly
  dominated to be confused with stars in the SDSS images (and for this reason
  they might not be included in the MPA-JHU galaxies catalog). However, this
  possibility can be discarded. First of all, in the Roma-BZCAT catalog of BL
  Lacs \citep{massaro09} there are only 8 objects with 0.05$\le z \le$ 0.1 and
  covered by the DR7. All of them have been selected as SDSS spectroscopic
  targets. Furthermore, the brightest of these objects has a radio luminosity
  of $1.3\times 10^{24}$ W Hz$^{-1}$. We estimated the corresponding non
  thermal contribution to the optical ($r$ band) magnitude for similarly
  bright objects by assuming, conservatively, a radio-to-optical spectral
  index of 0.5, typical of high energy peaked BL Lacs \citep{padovani03}. We
  obtain $r=17.5$ at $z=0.05$ (and $r=$19.0 at $z=0.1$), $\sim 1 $--4
  magnitudes fainter than LRP AGN hosts. Only much brighter BL~Lacs might be
  confused with stars. However, very luminous sources have a low space
  density and indeed no such object, that would have been easily discovered,
  is known within the volume considered. 

We conclude that our estimate of 267 objects is a robust upper limit to the
BL~Lacs number in our volume.

\begin{figure}
   \centering
   \includegraphics[width=87mm]{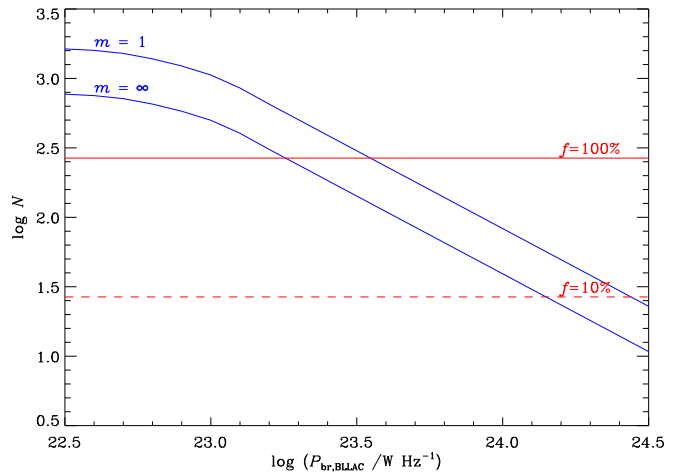}
   \caption{Total number of predicted BL~Lacs in the range
     $0.05 \le z \le 0.1$ in the SDSS/DR7 area as a function of the low
     luminosity break of their RLF, $P_{\rm br,BLLAC}$, and of the slope $m$ of
     the number density below the break. The solid horizontal line represents
     the total number of radio galaxies (weak lined and core dominated, see
     text for details) in the same volume; the dashed horizontal line
     corresponds to a BL~Lacs percentage of the selected radio-sources of
     10\%.}
    \label{bllac}
    \end{figure}

\citet{pad07} derived the RLF of BL~Lacs at 5 GHz based on the Deep X-ray
Radio Blazar Survey. They obtained a RLF consistent with a single power-law
with a slope of 2.12 $\pm$ 0.16 in the range $\log (P_{r} / \WHz) \sim 23.8
$--$ 26.8$.  By adopting this RLF we estimate the predicted number of BL~Lacs
in the selected volume when varying the luminosity, $P_{\rm br,BLLAC}$, at
which the RLF breaks, see Fig.\ \ref{bllac}. In our calculation we assume a
null radio spectral index\footnote{This assumption is based on the average
  radio spectral index of BL~Lacs with $z<0.1$, i.e. $<\alpha_{\rm r}> = 0.03$
  \citep{stickel91}.  A steeper radio spectrum would slightly increase the
  derived limits on $P_{\rm br,BLLAC}$.} to convert from 5 to 1.4 GHz and take
into account that objects with radio flux density smaller than 5 mJy are not
included in our sample (this effect causes the flattening of $\log N$ for
$\log (P_{\rm br,BLLAC} / \WHz)\lesssim 23$ in Fig.\ \ref{bllac}).  We
conclude that the number of BL~Lacs predicted by extrapolating their RLF to
low luminosities exceeds the total number of possible hosts for $\log (P_{\rm
  br,BLLAC} / \WHz) \lesssim 23.3 $--$ 23.5$, depending on the RLF slope below
the break. On the other hand, the RLF data points of \citeauthor{pad07}
exclude $\log (P_{\rm br,BLLAC} / \WHz) \gtrsim 24.5$.

We conclude that the RLF of BL~Lacs must break to a shallower slope for radio
luminosities smaller than $\log (P_{r} / \WHz)  \sim 23.3 $--$ 24.5$.

\section{Discussion and conclusions}

The results presented indicate that the RLFs at 1.4 GHz of both LRP AGN and
  BL~Lacs must decrease their slope at, or just above, the low luminosity
  limit at which they are currently determined. This might appear as a
  contrived coincidence, while, conversely, it is the natural consequence of
  the  behavior of the RLF at low power.  In fact, the number of objects
    below the break that can be detected in a flux limited survey depends on
    the radio power as $P_{\rm r}^{3/2+m}$, where $m$ is the number density
    slope for $P_{\rm r} < P_{\rm br}$. The strong dependence on $P_{\rm r}$
    implies that these objects are missing altogether in the currently
  available surveys (or at most found in a small number) and this prevents the
  RLF determination below $P_{\rm br}$.

\medskip
For LRP AGN we find that the number density below the break must significantly
decrease producing a peak in the distribution. It is tempting to associate the
presence of this peak to a minimal level of emission that is invariably
associated with a supermassive black hole located at the center of a massive
galaxy, which unavoidably accretes gas from the interstellar medium.  The peak
might instead correspond to a minimum level of accretion at which a radio jet
can be produced.  A better definition of the RLF at low luminosity is
essential in this context: in the first case, we expect to find that all
potential hosts harbor a LRP AGN, albeit of very low luminosity while, in the
second alternative, the occupation fraction would never reach 100\%. Note that
\citet{kimball11} found a similar dramatic fall of the number density from the
analysis of a sample of 179 QSOs ($M_i<$-23) from the SDSS in the redshift
range 0.2 $< z <$ 0.3, for radio luminosities below $\log (P_{r} / \WHz) \sim
21$ (at 6 GHz).

Unfortunately, we have only sparse information on the radio properties of
massive galaxies below $\log (P_{r} / \WHz) \sim 20$. These can be obtained
with deep targeted observations of nearby galaxies. In the Virgo clusters
there are two (out of 11) giant elliptical galaxies lacking of any sign of
radio emission down to a limit of $\log (P_{r} / \WHz) \sim 18.6$
\citep{capetti09}, despite their large mass ($M_* \sim 10^{11.5} M_\odot$) and
large black hole mass ($M_{\rm BH} \sim 10^{8} $--$10^{8.5} M_\odot$). The
small number statistics prevent us to derive any firm conclusion based on
these observations, but they apparently favor the interpretation that not all
massive galaxies are able to produce a radio jet.

\medskip
The two RLFs considered in this Letter are expected to be linked with each
other. Indeed, according to the AGN unified scheme, BL~Lacs are the beamed
version of low luminosity AGN \citep[see e.g.][]{urry95,tadhunter08}, i.e.,
BL~Lacs are objects in which a highly relativistic jet is seen at an angle
close to our line of sight. This accounts for their extreme properties, such
as strong flux variability and apparent superluminal motion of radio
components. The effects of relativistic beaming on the RLF was analyzed in a
series of papers \citep{urry84,urry91,urry91b,urry95}. If the RLF of the
parent population of unbeamed objects is described by a broken power law, the
break will produce a change of slope in the BL~Lac RLF at $\sim \delta^3_{\rm
  max} \times P_{\rm br,LRP}$, where $\delta_{\rm max}$ is the maximum value
of the jet Doppler factor distribution\footnote{The Doppler factor $\delta$ is
  defined as $\delta=[\Gamma (1-\beta \, \cos \theta)]^{-1}$, where
  $\Gamma=(1-\beta^2)^{-1/2}$ is the bulk Lorentz factor, $\beta$ is the
  velocity in units of the speed of light, and $\theta$ is the viewing
  angle. The exact value of the $\delta$ exponent depends on the geometry of
  the emitting region.}. The ratio between $P_{\rm br,LRP}$ and $P_{\rm
  br,BLLAC}$ is a factor $\sim 10^3$, which suggests a value of the Doppler
factor $\delta_{\rm max} \sim 10$, remarkably consistent with the typical
results obtained from the observations and the model predictions
\citep[e.g.][]{ghisellini93,savolainen10}. This analysis implicitly assumes
that all LRP AGN produce relativistic jets, a requirement that is not
necessarily met by the objects at the very faint end of the RLF. The
consistency between the estimates of the Doppler factor we just obtained with
the theoretical expectations appears to confirm the overall validity of this
assumption.

\medskip
How can we improve our knowledge of the RLF at low power? For LRP AGN the
lower limit to the break power, $\log (P_{\rm br,LRP} / \WHz) \sim 20.5$, is not far from what is currently accessible to radio observations
of large sample of galaxies such as the NVSS. A factor of $\sim$ 10
improvement in the flux threshold is already accessible to the JVLA and it
will be sufficient to measure directly the RLF of LRP AGN below the break
value we predicted. While waiting for such a survey it is possible to take
advantage of the stacking technique (e.g. \citealt{white07}) to measure the
median radio luminosity of the population of massive elliptical galaxies at
sub-mJy levels. In case the number density presents a peak in its
distribution, as our results suggest, this median luminosity is predicted to
be located close to the peak value.

\end{document}